\documentclass[usegraphicx]{mn2e}

\usepackage{mathptmx}
\usepackage{amssymb}
\usepackage{amsbsy}
\usepackage{url}
\usepackage[fleqn]{amsmath}

\begin{document}
\title{Resonance scattering, absorption and off-centre abundance peaks
  in clusters of galaxies}

\author[J.S.  Sanders and A.C. Fabian]{J.S.
  Sanders\thanks{E-mail: jss@ast.cam.ac.uk} and
  A.C. Fabian\\
  Institute of Astronomy, Madingley Road, Cambridge. CB3 0HA}
\maketitle

\begin{abstract}
  A possible explanation for the central abundance dips found from
  spatially-resolved X-ray spectroscopy of several groups and clusters
  of galaxies is resonance scattering. A number of the prominent iron
  emission lines are resonance lines. We construct a unique spectral
  model which takes account of resonance scattering for several
  thousand resonance lines, projection effects, photoelectric
  absorption, and allows direct spectral fitting.  We apply our model
  to \emph{Chandra} observations of two clusters with pronounced
  central abundance dips, Centaurus and Abell~2199. The results show
  that the effect of resonance scattering on emission from the centre
  of the cluster can be as much as 30~per~cent for the Fe-K resonance
  lines, and 10~per~cent for several Fe-L lines, if turbulence is low.
  The change to the metallicities obtained by fitting low resolution
  CCD spectra is at most 10~per~cent. Accounting for resonance
  scattering does not remove the central dip. Allowing for internal
  absorption within the Centaurus significantly improves the quality
  of the fits, indicating the presence of absorbing material within
  the inner 40~kpc of NGC~4696.
\end{abstract}


\section{Introduction}
Clusters traditionally known as cooling flow clusters, with highly
peaked surface brightness profiles and cool cores, typically host a
large galaxy in their very centre (Fabian 1994). These galaxies would
be expected to pollute their environment with metals, leading to
peaked abundance distributions. Such profiles are often seen in X-ray
observations of the intracluster medium (ICM) in cooling flow clusters
(e.g. Fukazawa et al 1994; De Grandi \& Molendi 2001).  However some
clusters and elliptical galaxies exhibit a dip in abundance in their
very centres, including Centaurus (Sanders \& Fabian 2002), Abell~2199
(Johnstone et al 2002), Perseus (Sanders et al 2004; Schmidt, Fabian
\& Sanders 2002; Churazov et al 2003) and NGC~4636 (Jones et al 2002).

Some possible reasons for these abundance dips are resonance
scattering (Gilfanov, Syunyaev \& Churazov 1987), incomplete spectral
modelling (e.g.  Buote 2000), a bimodel metallicity distribution
(Morris \& Fabian 2003), and complex flows (Mathews et al 2003; Fabian
2003).

Resonance scattering in clusters of galaxies was first considered by
Gilfanov et al (1987). They pointed out that clusters of galaxies
could be optically thick at the energies of strong resonance lines.
Resonance transitions are transitions from the ground state of an ion.
They are more likely to occur than other transitions as there are a
large population of ions in the ground state. If there is a sufficient
ion column density along a line-of-sight, radiation which would
otherwise travel to the observer is scattered out of that line-of-sight
into another one. As many resonance lines are strong emission lines,
this may lead to an underestimate of the abundance of the gas in the
centre of the cluster, and an overestimate in the outer parts of the
cluster.

We describe here a spectral model which includes the effects of
resonance scattering, for use in fitting X-ray spectra from clusters
of galaxies. By applying this model to data from clusters exhibiting
an off-centre abundance peak we will assess whether resonance
scattering can explain these effects.  Some authors report that
resonance scattering is not important for the bulk of the X-ray
emission from the core of the Perseus cluster (Churazov et al 2004;
Gastaldello \& Molendi 2004). It is however still unclear whether the
emission from the very centre of the cluster, or the emission in
L-shell lines are significantly affected by resonance scattering.

Several other authors have modelled resonance scattering in clusters,
including a Monte Carlo radiative transfer method used on data from
M87 (Mathews, Buote \& Brighenti 2001), and a radiative transfer
method by Shigeyama (1998) in elliptical galaxies.  It should also be
stated that resonance scattering has been considered important when
modelling the solar corona (e.g. Acton 1978).

Rather than a full radiative transfer code, we have aimed for a
simpler method which can be calculated in real-time whilst fitting
X-ray spectra using \textsc{xspec} (Arnaud 1996). We assume the
single-scattering approximation, and that photons are conserved.  The
method calculates the emission and scattering as a function of
spectral energy in shells in the cluster. By integrating along
lines-of-sight, model X-ray spectra from annuli on the sky are
calculated for a particular temperature, abundance and density cluster
profile.  The model spectra are fit against data extracted from the
appropriate annuli on the sky.

The model can also self consistently account for photoelectrically
absorbing material within the cluster. It is interesting to study the
amount of this material as it is a possible contributing factor to the
lack of emission lines seen emitted from cooling gas (e.g. Fabian et
al 2001), although such absorption would need to operate
preferentially on the emission from the cool gas.

We assume the solar abundance ratios of Grevesse \& Sauval (1998). For
distance calculation, we assume that $H_0 = 70
\;\mathrm{km}\;\mathrm{s}^{-1}\;\mathrm{Mpc}^{-1}$. All uncertainties
shown are 1-$\sigma$.

\section{Model}
For a particular annulus, the model emission spectrum is calculated by
summing the emission from each spherical shell along its line of sight
through the cluster. If there are $N$ annuli on the sky, then $N$
shells are used to model the cluster. Each shell is divided into $n_s$
``subshells'' to allow for the variation of gas properties across the
shell. A typical value of $n_s$ we use is 3. In addition each annulus
on the sky is similarly divided into $n_s$ ``subannuli''.  The
spectrum from an annulus is the sum of the spectra from its subannuli.

The electron density, temperature and solar relative abundance are
specified at the centre of each shell. The gas properties of subshells
are found using linear interpolation in logarithmic radius of the
properties from the centres of the nearest shells, or by just using a
fixed value within a shell. We later find that using constant values
within each shell appears to produce more consistent results.

\begin{figure}
  \includegraphics[width=\columnwidth]{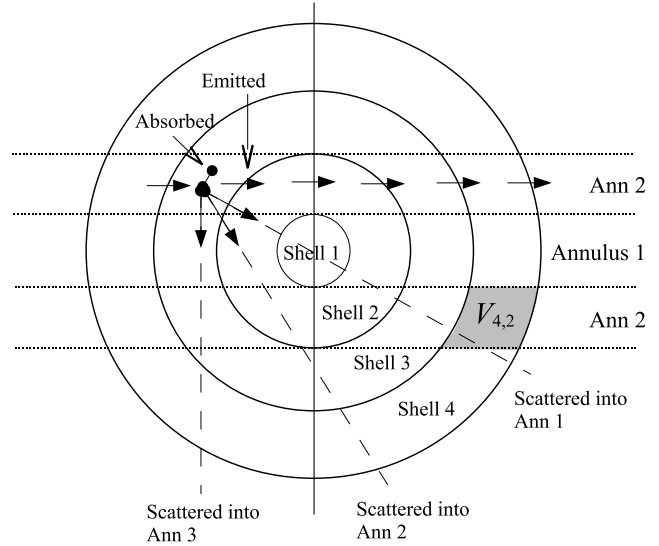}
  \caption{Diagram showing schematics of scattering modelling. This is
    a cross section of the cluster along the line of sight, with the
    observer to the right of the diagram. In this example we are
    looking at the emission summed along the line of sight of annulus
    2 (subannuli and subshells are ignored for clarity here).}
  \label{fig:schem}
\end{figure}

In general, the model works as follows (Fig.~\ref{fig:schem}). We
iterate over each of the subannuli on the sky. For a particular
subannulus, we start at the far side of the cluster. We calculate the
emission spectrum from the volume of intersection of the outermost
subshell with that subannulus. This spectrum is carried forward
towards the observer into the volume of intersection with the next
subshell. Using the list of resonance lines, we compute how much
radiation is scattered as a function of energy in that volume, and
remove this radiation from the summed spectrum. The thermal emission
from the volume is added to the sum, and the calculation carries on to
the next subshell.

To account for radiation scattered into other lines of sight, we keep
track of the total radiation scattered by a particular subshell. After
we have summed along each line of sight, this spectrum is then
distributed among the subannuli according to the fraction expected
from Rayleigh scattering.

In addition the model can also take account of X-ray photoelectric
absorption within each subshell, applying it consistently. Within each
volume of intersection, radiation is removed according to the column
density the radiation intercepts. Absorption is also applied to
radiation scattered into other lines of sight.

Since the model cluster is spherically symmetric, it is only necessary
to do the calculation along one line of sight.  We discuss each of the
steps of the model in detail below.

\subsection{Determination of resonance lines}
We used the \textsc{aped} atomic data database (version 1.3.1; Smith
et al 2001a) to generate a list of resonance lines to use in the
model. This enabled us to use the same source of lines as was used to
calculate the emission spectra (\textsc{apec}; Smith et al 2001b).

For each ion, we identified those lines involving transitions from the
ground state (i.e.  the resonance lines). We took their Einstein-A
coefficients, $A_{ug}$, wavelengths, $\lambda$, and upper and lower
statistical weights for the energy levels the transition is between,
$g_u$ and $g_g$. From these we calculated the oscillator strength of
the lines, which in SI units is for each
\begin{equation}
  f_{ug} = -\frac{A_{ug}}{3 \gamma} \frac{g_u}{g_g}
\end{equation}
where if $\omega_0 = 2 \pi c / \lambda$,
\begin{equation}
  \gamma = \frac{1}{4 \pi \epsilon_0} \frac{2}{3}
  \frac{e^2}{m c^3} \omega_0^2.
\end{equation}

\begin{figure}
  \includegraphics[width=\columnwidth]{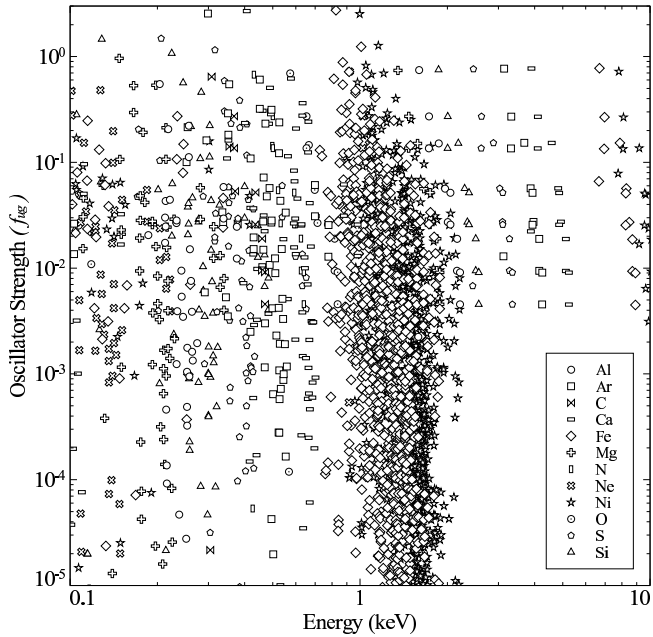}
  \caption{Plot of all the resonance lines above 0.1~keV in
    \textsc{aped} with an oscillator strength of greater than
    $10^{-5}$.}
  \label{fig:reslines}
\end{figure}

In Fig.~\ref{fig:reslines} we show the oscillator strengths of the
resonance lines above 0.1~keV and $f_{ug} > 10^{-5}$, identified by
element. The relative importance of particular lines will vary with
oscillator strength and the abundance of the particular ion, which is
a function of temperature. Our model includes all 3574 resonance lines
in the \textsc{aped} database, with no oscillator-strength lower
threshold.

\subsection{Calculation of the emission spectra}
\label{sect:emissspec}
If we are fitting spectra from $N$ annuli on the sky, with spectrum
$a$ extracted from radii on the sky $\Pi_{a-1}$ to $\Pi_{a}$, then we
fit for the temperature, $kT_a$, the abundance, $Z_a$, and the
electron density, $n_{e,a}$ at a radius $(\Pi_{a-1}+\Pi_{a})/2$ in the
cluster, assuming spherical symmetry.

\begin{figure}
  \includegraphics[width=\columnwidth]{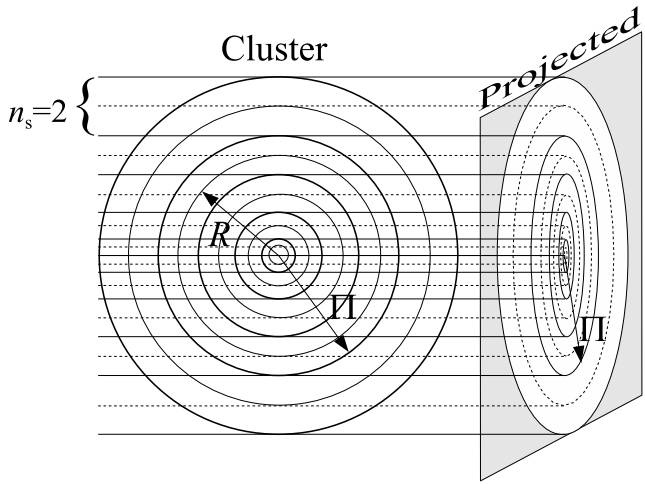}
  \caption{The mapping of the shells and subshells in the cluster, to
    the annuli and subannuli on the sky. In this example we have 2
    subshells and subannuli.  Radius $\Pi$ corresponds to the radii of
    the annuli on the sky and the shells in the cluster, and radius
    $R$ corresponds to the radii of the subshells in the cluster.}
  \label{fig:projdia}
\end{figure}

The cluster is modelled by shells in the cluster corresponding to the
annuli on the sky. Each shell is divided into $n_\mathrm{s}$ subshells
(Fig.~\ref{fig:projdia}). This gives a total of $Nn_\mathrm{s}$
subshells, where $n_\mathrm{s}$ is chosen to be as large as possible
without making the calculation too slow. Subshell $a$ lies between a
radius of $R_{a-1}$ and $R_a$ in the cluster. We use either a fixed
temperature, abundance and density in the subshells within each shell,
or linear interpolation in logarithmic radius of these values from the
centre of each shell.  The second option allows for a smooth
distribution of gas components between each shell, which we refer to
as ``interpolation'' in this paper. When interpolating, the properties
within the inner half of the inner shell and the outer half of the
outer shell are extrapolated in logarithmic radius from the
neighbouring points.

For each subshell, $a$, in the set of shells we calculate the spectrum
of the emitting plasma per unit volume, $\pmb{S}_a$, at high spectral
resolution in order to adequately sample the major spectral lines over
their line width. We used $10^5$ linear spectral elements between 0.3
and 9 keV, giving a spectral resolution of 0.087~eV. This is around
half the thermal width of a 0.6~keV oxygen line at temperatures of
0.6~keV, or a 1~keV iron line at the same temperature.

The spectrum is calculated using a rewritten implementation of the
\textsc{apec} model in \textsc{xspec}, highly optimised for the case
of fixed spectral resolution.  It works by precalculating
high-resolution spectra at the \textsc{apec} grid-points.  The
implementation also includes a Gaussian approximation for the thermal
widths of the spectral lines plus an additional turbulence component.
The total Doppler width used is (e.g. Mathews et al 2001)
\begin{equation}
  \Delta \nu_D = \frac{\nu_0}{c} \left(
    \frac{2kT}{A_{Ewt} m_p} + v^2_\mathrm{turb} \right)^{1/2},
\end{equation}
where $\nu_0 = c/\lambda$ is the central line frequency, $A_{Ewt}$ is
the atomic weight of the element producing the line, $m_p$ is the
proton mass, and $v_\mathrm{turb}$ is the turbulent velocity. The line
shape is calculated out to three times $\Delta \nu_D$ either side of
the central frequency.

\subsection{Scattering calculation}
In addition to the emission spectrum in each subshell, also calculated
is the scattering coefficient, $\pmb{\kappa}_a$, in units of
length${}^{-1}$, over the same energy range and resolution as the
emitted spectrum. In order to calculate the scattering coefficient we
iterate over each of the resonance lines in the energy range
considered. For each line we calculate the scattering profile using
(again Mathews et al 2001)
\begin{equation}
  \kappa_\mathrm{line}(x) = n_H \: x_\mathrm{ion} \: a_E \: A_E
  \left( \frac{e^2}{4 \epsilon_0 m_e c}
    f_{gu}
    \frac{1}{\pi^{1/2}}
  \frac{1}{\Delta \nu_D}\right) e^{-x^2},
\end{equation}
where $n_H$ is the hydrogen density of the shell, found using $n_H =
n_e / 1.2$, $A_E$ is the abundance of the element producing the line
relative to solar, $a_e$ is the solar abundance of the element
relative to Hydrogen, and $x_\mathrm{ion}$ is the fraction of the
element ionised into the ion producing the line (which is a function
of the temperature of the subshell). We use the ionisation fractions
of Mazzotta et al (1998), interpolating between the provided values.
The dimensionless line frequency, $x=(\nu-\nu_0)/\Delta \nu_D$.

We add the line scattering profile (calculated out to three times
$\Delta \nu_D$ from the central frequency) to the total `scattering
coefficient spectrum' for the subshell, accounting for redshift.

\subsection{Calculating the observed spectrum}
To model the spectrum in an annulus, we divide the annulus into
$n_\mathrm{s}$ subannuli (this is chosen to be the same as
$n_\mathrm{s}$ as was used to create the subshells in Section
\ref{sect:emissspec}, but this need not be the same value). Along the
centre of each subannulus, we integrate along the line-of-sight from
the back of the cluster to the observer. We take account of the
radiation emitted in a particular subshell, and the radiation
scattered out of the line-of-sight by that subshell. Optionally,
absorbed radiation is also removed. When the spectra for all of the
subannuli has been calculated, the code takes the radiation scattered
by a particular subshell, computes the fraction scattered into each
annulus on the sky, and adds this back to give the final model
spectra. Absorption is also optionally applied.

The model for a particular annulus is then computed by summing the
integrated spectrum of its subannuli with appropriate area factors.
\textsc{xspec} fits the model by minimising the sum of the $\chi^2$ or
C-statistic of the fits to the individual input spectra.

In detail, the procedure for calculating the emission along each
line-of-sight is below, representing spectra as vectors in bold. For
subannulus $a$ we sum the emission along its midpoint in radius on the
sky, $R_{mid,a} = (R_a + R_{a-1})/2$.

\begin{enumerate}
\item We start with a blank output spectrum, $\pmb{O}_a \leftarrow
  \mathbf{0}$, and consider subshell $b=N n_s$, where the
  line-of-sight intercepts the cluster on the outer side away from the
  observer (Fig.~\ref{fig:schem}).
\item \label{item:iterate} Take the `scattering spectrum' for the
  subshell, $\pmb{\kappa}_b$, multiply by the line-of-sight length,
  $l_b$, between subshell $b$ and $b-1$ at radius on the sky
  $R_{mid,b}$, to make an `optical depth spectrum', $\pmb{\tau}_b$.
  To estimate an effective value of $l_b$ we take half the volume
  between subshell $b$ and $b-1$ intercepting the lines-of-sight
  between radii on the sky $R_a$ and $R_{a-1}$, $V_{ba}$
  (Fig.~\ref{fig:schem}), and divide by the area on the sky of the
  subannulus, $A_a = \pi (R_a^2 - R_{a-1}^2)$. $V_{ba}$ is given by
  the integral, 
  \begin{equation}
    V_{ba} =
    {\displaystyle \int}_{y=R_{a-1}}^{R_a}
    \left( \sqrt{R_{b}^2 - y^2} - \sqrt{R_{b-1}^2 - y^2} \right)
     2 \pi y \: \mathrm{d}y.
  \end{equation}
  To allow for the fact that the annuli may on the sky may have larger
  radii than the shells in the cluster, we truncate the integration to
  where the square-roots are real. This leads to an expression for the
  volume of
  \begin{align}
    V_{ba} = \frac{2}{3} \pi \Big[&
    Q( R_{b-1}^2 - R_a^2 ) - Q( R_{b-1}^2 - R_{a-1}^2 )
    \nonumber \\
    & + Q( R_b^2 - R_{a-1}^2 ) - Q( R_b^2 - R_a^2 )
    \Big],
  \end{align}
  where
  \begin{equation}
    Q(x) = \left\{ \begin{array}
        {r@{\quad:\quad}l}
        x^{3/2} & x \ge 0 \\
    0 & x < 0.
  \end{array} \right.
\end{equation}

We multiply $\pmb{O}_a$ at each spectral energy by
$\mathrm{exp}(-\pmb{\tau})$ at each spectral energy to take account of
the scattering out of the line-of-sight in $b$,
\begin{equation}
  \pmb{O}_{a,i} \leftarrow \pmb{O}_{a,i} \times \mathrm{exp}
  \left[ -\frac{V_{ba}}{A_a} \pmb{\kappa}_{b,i} \right]
  \textrm{ for each energy bin, } i.
\end{equation}
To take account of this radiation being scattered into other
lines-of-sight, we keep track of a total spectrum of the radiation
which is scattered in subshell $b$.

\item Optionally, absorption within the shell is included in the
  model. $\pmb{O}_a$ is multiplied with the result of a \textsc{phabs}
  absorber of column density $N_\mathrm{H}$, where $N_\mathrm{H} =
  n_\mathrm{H} l_b$, if $n_\mathrm{H}$ is the equivalent number of
  absorbing hydrogen atoms per unit volume, assuming solar abundances.

  In detail, the model precalculates the optical depth for unit
  absorption to improve its speed. In addition the absorption spectrum
  is stored at lower energy resolution than the rest of the model (we
  used 8.7~eV energy resolution). This greatly improves the speed of
  the model. The effect of this simplification to the output spectrum
  is minimal as the energy dependence of the absorption is smooth,
  away from the few edges in the spectrum.

\item To take account of the emission, we take the emission spectrum
  per unit volume for subshell $b$, $\pmb{S}_b$, multiply by $V_{ba}$,
  and add to the output spectrum
\begin{equation}
  \pmb{O}_{a,i} \leftarrow \pmb{O}_{a,i} + V_{ba} \: \pmb{S}_{b}
  \textrm{ for each } i.
\end{equation}

\item Move to the next innermost subshell, i.e. $b \leftarrow b-1$,
  and go back to \ref{item:iterate}. When $b$ is the innermost
  subshell, iterate over that subshell again, then iterate over
  increasing $b$ until $b=N n_s$. This means that we integrate over
  the shells on the side of the cluster away from the observer,
  through the core of the cluster, and out towards the observer to the
  edge of the cluster again.
\end{enumerate}

The model spectrum for an annulus $a$ is the sum of
$\pmb{O}_{n_\mathrm{s} (a-1)}$ to $\pmb{O}_{n_\mathrm{s}a - 1}$,
plus the contribution of radiation scattered into the line-of-sight.

To calculate the radiation scattered into a line-of-sight, we first
take the spectrum of the accumulated radiation which is scattered out
of all the lines-of-sight by a particular shell. That spectrum is then
divided among the annuli on the sky according to the fraction of
radiation calculated to be scattered into each (see
Appendix~\ref{appen:scat}). This procedure conserves photon flux, but
does not include the physics of the broadening of lines in the
scattering process. This effect is irrelevant in the single-scattering
assumption unless one is using an instrument which could measure line
widths. It is difficult to exactly account for this effect in the
current model.

The effects of absorption on the scattered radiation can optionally be
accounted for. The model sums up the contribution to the scattered
radiation by considering the shells which could scatter into a
particular subannulus. Starting from the far side of the cluster, the
contribution to the scattered spectrum is calculated for the outermost
half-subshell. This radiation is carried to the half-subshell next
nearest the observer, absorption is applied within this subshell, and
the scattered contribution added. We repeat this process until we
reach the half-subshell nearest the observer.

After generating the high resolution model spectra, they are
downsampled to instrument resolution for spectral fitting in
\textsc{xspec}.

\section{Application of model}
\subsection{Centaurus cluster}
We applied our model to the Centaurus cluster which has a clear
abundance drop in its centre (Sanders \& Fabian 2002; Fabian et al
2005). The details of the observations analysed are shown in
Table.~\ref{tab:obs}, and are the same observations as discussed in
Fabian et al (2005). Each of the observations were take using the
ACIS-S3 back-illuminated CCD. The \textsc{lc\_clean} tool was used to
filter periods with flares, using the lightcurve of the ACIS-S1
back-illuminated CCD in the band 2.5 to 7~keV, yielding a total
exposure of 196.6~ks.  Each of level-1 event files for the
observations were reprocessed using the PSU CTI corrector (Penn State
University corrector for charge transfer efficiency), version 1.45,
(Townsley et al 2002a,b). Blank sky background files were used during
spectral fitting. The exposure times of the background event files
were modified to ensure the same count rate as the observations in the
9 to 12~keV band (where background dominates over source counts). We
used a response matrix generated using the PSU ACIS CCD simulator,
using an energy resolution of 4~eV. Ancillary response matrices were
created using the \textsc{ciao} \textsc{mkwarf} tool (\textsc{ciao}
version 3.2.2) and the PSU QEU file, weighting the spatial regions
using the 0.5 to 7.0 keV band. Background spectra were extracted from
the same spatial region in blank-sky event files.

\begin{table}
  \caption{\emph{Chandra} observations of Centaurus analysed with
    clean exposure time.}
  \begin{tabular}{lll}
    Observation & Observation date & Clean exposure (ks) \\ \hline
    504  & 2000-05-22 & 21.8 \\
    505  & 2000-06-08 & 10.0 \\
    4954 & 2004-04-01 & 80.2 \\
    4955 & 2004-04-02 & 40.6 \\
    5310 & 2004-04-04 & 44.1 \\ \hline
    Total &           & 196.6 \\
  \end{tabular}
  \label{tab:obs}
\end{table}

\begin{figure}
  \includegraphics[width=\columnwidth]{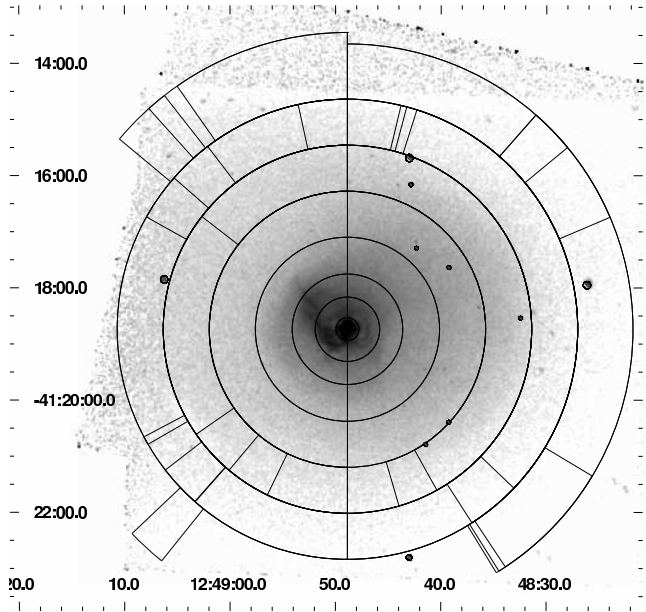}
  \caption{Regions used for spectral fitting. The image is exposure
    map corrected, and smoothed by a Gaussian of 1~arcsec. 
    The western half of the cluster is least disturbed in the central
    regions.} 
  \label{fig:regions}
\end{figure}

We analysed the cluster in two spatial parts. Our previous work showed
that the eastern and western halves are distinct. There is a
plume-like feature in the eastern central regions. Therefore we
concentrated on the western half, where the central regions show less
disturbance.  Fig.~\ref{fig:regions} shows the regions used in this
analysis. As the different datasets examined were taken with different
roll-angles and aim-points, the starting and stopping angles of the
sectors were modified for each dataset to fit on the S3 CCD.

We fit the spectra in \textsc{xspec} using our resonance scattering
model, fitting the datasets simultaneously, and minimising the
$\chi^2$ during the fit. The spectra were fit between 0.5 and 7~keV,
with the spectra grouped to have a minimum of 20 counts per spectral
bin. A \textsc{phabs} (Balucinska-Church \& McCammon 1992) component
was used to account for Galactic absorption, applied equally to each
annulus. The absorption column density of this component was allowed
to be free in the fits.

In Fig.~\ref{fig:cenprof} we show the results of the spectral fits for
the western and eastern half of the cluster using a variety of
variants on the resonance scattering model. The points plotted show
the model optionally including the effects of resonance scattering,
interpolation of the gas properties to smooth the profile, and fitting
for internal absorption within the cluster. The quality of the fits
are shown in Table~\ref{tab:cenfits}. We also calculated the profiles
using radial bins with half the width in the Western half of the
cluster (Fig.~\ref{fig:cenproffine}).

\begin{figure*}
  \includegraphics[width=\columnwidth]{fig5_left.fix.epsi}
  \hspace{2mm}
  \includegraphics[width=\columnwidth]{fig5_right.fix.epsi}
  \caption{Radial electron density (top), temperature, abundance, and
    internal photoelectric absorption (bottom) profiles for the west
    (left) and east (right) of the Centaurus cluster.  Each symbol and
    line shows the profile with a variant of the model. ``Reson.''
    indicates a model including the effects of resonance scattering
    and ``interpol.''  uses a smooth range of components (5 in west
    and 3 in east) within each shell.  ``absorb.''  indicates that
    internal absorption is fitted for within the cluster. The radius
    marked is the mean radius of each shell. The points have been
    offset from the black circle slightly horizontally for clarity.}
  \label{fig:cenprof}
\end{figure*}

\begin{table*}
  \caption{Quality of fits for the various models and data for
    Centaurus. The sectors used are West, the West with finer
    sectors (half the width of the normal sectors), and East. If resonance
    scattering is indicated, then resonance scattering was modelled
    (with no turbulence). Interpolation
    specifies whether the properties of the gas were interpolated
    within the subshells (5 for thick sectors, west, 3 for others).
    The absorption column indicates whether
    internal photoelectric absorption was modelled within the
    cluster.}
  \begin{tabular}{lllllll}
    Sector & Resonance scattering & Interpolation & Absorption &
    $\chi^2$ & dof & Reduced $\chi^2$ \\ \hline

    West & No  & No  & No  & 13039 & 9533 & 1.37 \\
    West & Yes & No  & No  & 13005 & 9533 & 1.36 \\
    West & No  & Yes & No  & 12938 & 9533 & 1.36 \\
    West & Yes & Yes & No  & 12868 & 9533 & 1.35 \\
    West & No  & No  & Yes & 12613 & 9525 & 1.32 \\
    West & No  & Yes & Yes & 12511 & 9525 & 1.31 \\
    West & Yes & No  & Yes & 12626 & 9525 & 1.33 \\
    West & Yes & Yes & Yes & 12417 & 9525 & 1.30 \\
    West (fine) & No  & No  & No  & 18735 & 15252 & 1.23 \\
    West (fine) & Yes & No  & No  & 18651 & 15252 & 1.22 \\
    West (fine) & No  & Yes & No  & 18591 & 15252 & 1.22 \\ \hline
    East & No  & No  & No  & 13525 & 8774 & 1.54 \\
    East & Yes & No  & No  & 13439 & 8774 & 1.53 \\
    East & No  & Yes & No  & 12989 & 8774 & 1.48 \\
    East & Yes & Yes & No  & 12797 & 8774 & 1.46 \\
    East & No  & No  & Yes & 12576 & 8766 & 1.43 \\
    East & No  & Yes & Yes & 12503 & 8766 & 1.43 \\
    East & Yes & No  & Yes & 12700 & 8766 & 1.45 \\ 
    East & Yes & Yes & Yes & 12509 & 8766 & 1.43 \\
  \end{tabular}
  \label{tab:cenfits}
\end{table*}

\begin{figure}
  \includegraphics[width=\columnwidth]{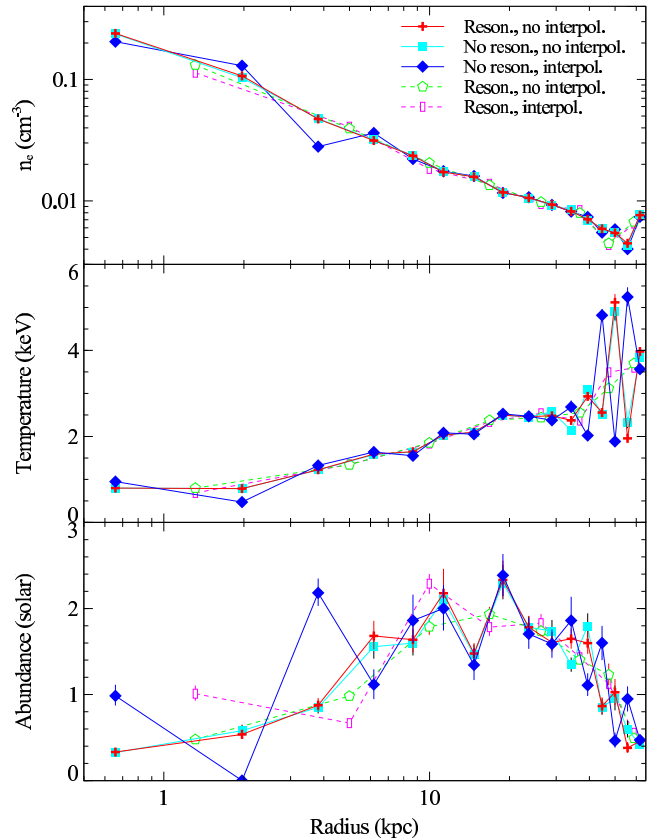}
  \caption{Profiles for the Western half of Centaurus using radial
    bins with half the width. Overplotted in hollow symbols are
    results from Fig.~\ref{fig:cenprof} including and not including
    interpolation of the gas properties.}
  \label{fig:cenproffine}
\end{figure}

Accounting for resonance scattering improves the quality of the
spectral fits slightly. Allowing multiple interpolated components per
shell and allowing for internal absorption improves the fits much
more.

It can be seen, as we will discuss later, that the effect on the
derived abundance including resonance scattering is generally small.
Greater changes to the metallicities are found using models which
interpolate the gas properties across each shell. However the
interpolation feature seems to exacerbate instabilities in the
spectral fit. This is particularly evident in the outer regions of the
temperature profile of the eastern half.  To gauge whether the
interpolation technique improves the accuracy of the results, the
lower spatial resolution interpolated and non-interpolated results can
be compared with those using higher spatial resolution
(Fig.~\ref{fig:cenproffine}). Those profiles from models which do not
interpolate the gas properties appear to be consistent. Interpolated
models produce inconsistent results, suggesting that the assumption of
spherical symmetry is inadequate, or there is additional non-projected
multiphase material.  The second most important factor appears to be
modelling absorption, which we discuss separately in
Section~\ref{sect:absorb}.

\begin{figure}
  \includegraphics[width=\columnwidth]{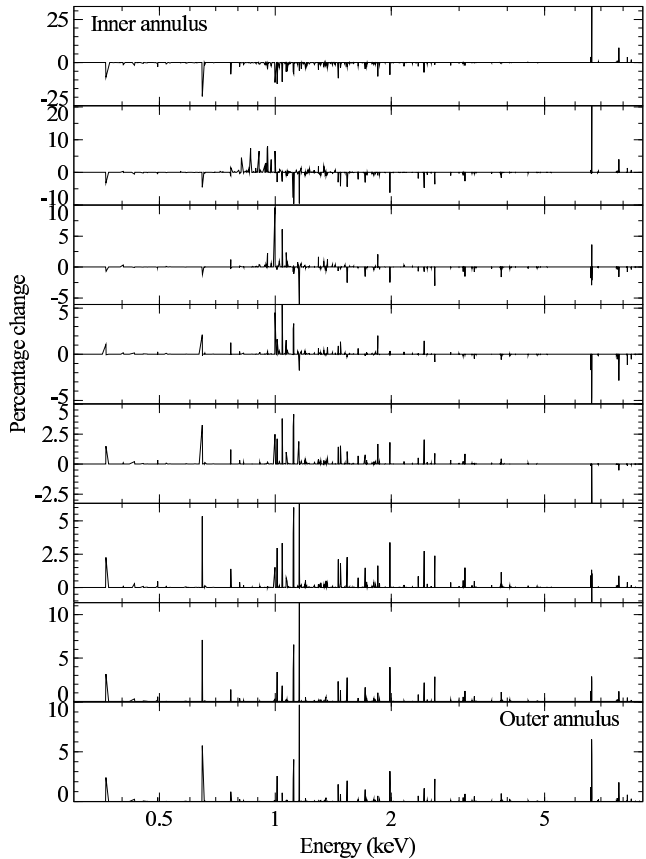}
  \caption{Fractional difference between the best fitting resonance
    scattering model spectrum for the western half of the Centaurus
    cluster (Fig.~\ref{fig:cenprof} left), and with the model with
    resonance scattering disabled. Here a model without absorption or
    interpolation was used.  A positive value shows more
    emission in the resonance scattering model. The top panel shows
    the results for the innermost shell, and the bottom the outermost
    shell. No turbulent velocity components are included here.}
\label{fig:cenratio}
\end{figure}

\begin{figure}
  \includegraphics[width=\columnwidth]{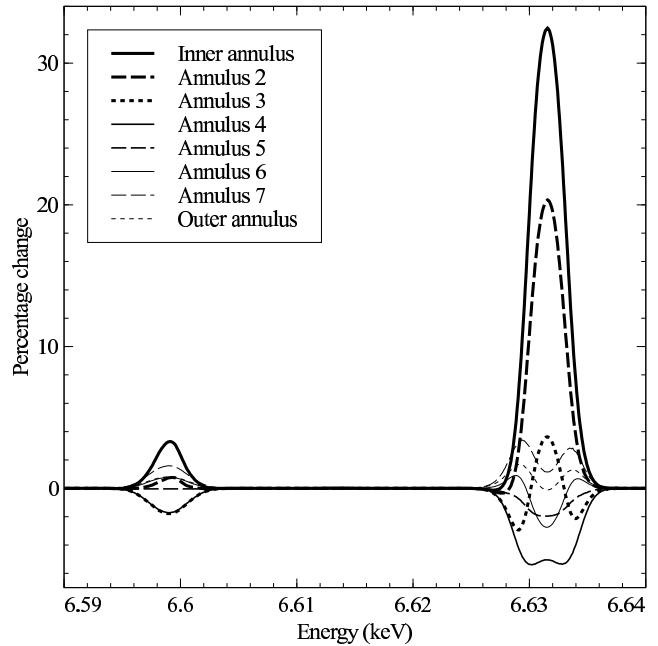}
  \caption{Detail of Fig.~\ref{fig:cenratio} around the Fe-K
    lines. Positive values means that the resonance scattering model
    predicts more emission from the shell than simple projection.}
  \label{fig:cenratiofek}
\end{figure}

To quantify the effect that the resonance scattering model is having
on the spectrum, we show in Fig.~\ref{fig:cenratio} the fractional
difference between the best fitting non-interpolated resonance
scattering model, and that model with resonance scattering disabled.
In the innermost shell the maximum effect on Fe-K lines is around
30~per~cent, and around 10 per~cent for the Fe-L lines. The overall
trend is what is naively expected: Fe-L is scattered out of the cooler
inner shells into outer shells, but Fe-K is scattered out of the
hotter outer shells into the inner shells (see
Fig.~\ref{fig:cenratiofek} for detail).

We also carried out spectral fitting with an additional
$300$~km~s${}^{-1}$ turbulent velocity component. The results from
these fits lay between the results without resonance scatter, and with
resonance scattering, as would be expected.

The optical depth of the scattering (ignoring radiation scattering
back into the line of sight) is around 1.3 for one of the Fe-K lines,
if we examine the radiation emitted from the centre of the cluster. A
few of the Fe-L lines approach optical depths of unity. The
probability of these photons being scattered again is in fact smaller
than a naive calculation would give, as the lines are broadened by the
scattering. The single scattering approximation is justified by this.

\subsection{Abell 2199}
We also applied our model to Abell~2199, which was found to have an
abundance drop by Johnstone et al (2002). A2199 is a rich cluster of
galaxies at a redshift of 0.0309. The cluster has a temperature of
4.7~keV (Edge \& Stewart 1991), and a peaked surface brightness
profile.

We examined spectra extracted from \emph{Chandra} observations 497 and
498, yielding a total exposure of 33.3~ks after removal of flares.
The same regions for spectral extraction as were used by Johnstone et
al (2002) for their results including projection effects, were chosen.
The event files were reprocessed with the latest appropriate gain
files in \textsc{caldb} 3.01. We used blank-sky fields for background
subtraction. We minimised the $\chi^2$ statistic during the spectral
fitting, and fitted the data between 0.5 and 7~keV. Again, we
accounted for Galactic absorption with a free \textsc{phabs}
component. We did not use the PSU CTI corrector on this dataset as the
one of the observations was taken before the date the corrector is
valid from.

\begin{figure}
  \centering
  \includegraphics[width=\columnwidth]{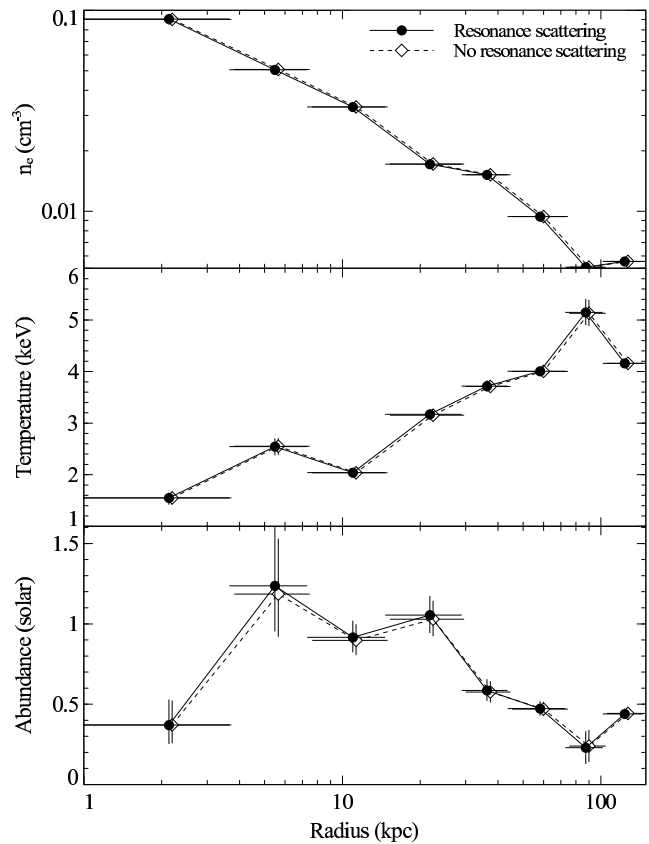}
  \caption{Density, temperature and abundance profiles for Abell 2199.
    Shown are the results including resonance scattering effects, and
    not including such effects. No subshell interpolation was used in
    these results. The results without resonance scatter have been
    offset horizontally for clarity.}
  \label{fig:a2199}
\end{figure}

In Fig.~\ref{fig:a2199} are shown the radial profiles from the results
of our spectral fitting, showing the results accounting for resonance
scattering and without. Again we see that the effect of including
resonance scattering in the model is small on the measured
abundances.

\begin{table}
  \caption{Effect on the quality of the spectral fit for A2199, when
    including resonance scattering. There is virtually
    no change to the quality of the fit, although the fit including
    resonance scattering is slightly better.}
  \begin{tabular}{lllll}
    Resonance scattering & $\chi^2$ & dof & Reduced $\chi^2$ \\ \hline
    No  & 4113.9 & 3694 & 1.11 \\
    Yes & 4105.8 & 3694 & 1.11 \\
  \end{tabular}
  \label{tab:chi2a2199}
\end{table}

\subsection{Absorption}
\label{sect:absorb}
If only Galactic absorption is assumed, there is evidence for low
energy residuals in the fits to the inner region of the eastern half
of the Centaurus cluster. This may be indicating that there is
photoelectrically absorbing material in that region. There is also
evidence for small spatial scale absorbing material in the central
regions of the cluster with a similar morphology to the dust lane in
NGC~4696 (Crawford et al 2005).  The Centaurus \emph{Chandra} data are
of such high quality that we are able to test for the presence of an
absorbing component within the cluster for both sectors. Allow for
absorption within each shell obtains measurements for the amount of
absorbing material per unit volume shown in the bottom panel of
Fig.~\ref{fig:cenprof}.

The amount of absorption found in the western half of Centaurus
appears fairly consistent between the model variants, except perhaps
in the very innermost annulus. The level of absorption appears to increase
inside 40~kpc radius, in the form of a rough power law. The region
with the apparent absorbing material appears to be only that gas below
around 2.5~keV. The minimum value of $0.08 \times 10^{20} \:
\mathrm{cm}^{-2} \: \mathrm{kpc}^{-1}$ within the inner 40~kpc, would
indicate an equivalent absorbing hydrogen density of $3 \times 10^{-3}
\: \mathrm{cm}^{-3}$. Within 40~kpc this would correspond to a mass of
$2 \times 10^{10} \: \mathrm{M}_\odot$ of absorbing material assuming
solar relative abundances.

We should note, however, that there is a build-up of contaminant on
the ACIS detector, which is not spatially uniform (see Sanders et al
2004 for a map of this contaminant from a Perseus cluster
observation). Although the \textsc{mkwarf} tool that we used to create
the ancillary responses includes a correction for the spatial
variation of this contaminant over time, there may still be
uncertainties in this calibration. It is possible that the evidence we
show is due to calibration uncertainties. The absorbing material on
small spatial scales found by Crawford et al (2005) indicates that the
calibration is not likely to be responsible for all of the signal
within the central regions. It will be important to confirm the
absorption detection in data from \emph{XMM-Newton} in future.

\section{Discussion}
\subsection{Resonance scattering}
Our model shows that the effect of resonance scattering in the two
clusters of galaxies examined is minimal to the determination of
abundances from CCD resolution X-ray spectra. The change in abundance
measured is less than 10~per~cent. The model assumes that spherical
symmetry is a good approximation in these clusters. It also ignores
the effects of non-Solar abundance ratios, or the presence of
multiphase gas which is not the result of projection.

The largest effect on the determined abundance in the central regions
is using a model which allows for an interpolated range of components
in each shell. Although this should produce results which more closely
match the real profile if the gas properties vary smoothly, it seems
that it is prone to effects which cause instability and oscillating
profiles (e.g. in the outer parts of the eastern temperature profile,
the western profile at high spatial resolution;
Fig.~\ref{fig:cenproffine}, and the inner region of the eastern
abundance profile). We suspect that non-spherical geometry may be a
the cause of this instability. The inner parts of the cluster show
obvious asymmetric features. The cluster, centred on NGC~4696, is also
thought to be merging or has merged with a subcluster 15~arcmin to the
east centred on NGC~4709 (Allen \& Fabian 1994; Churazov et al 1999;
Furusho et al 2001; Dupke \& Bregman 2001). The oscillations are
particularly evident in the outer regions on the eastern sector.

Although resonance scattering appears not to be important when
examining low spectral resolution spectra from clusters of galaxies
(at least in clusters similar to Centaurus), it is an important
process when examining high resolution X-ray cluster spectra in
detail. The effect increases the strength of the Fe-K resonance lines
by up to 30~per~cent from the centre of the Centaurus cluster if
turbulence can be neglected.  The effect on Fe-L resonance lines is
smaller ($< 10$~per~cent), but there are many more of them. When
fitting the low resolution CCD spectra from \emph{Chandra} the effect
on the obtained abundance is small ($< 10$~per~cent). However, if
turbulence is low the effect will be important to consider when
examining high spectral resolution data. The straightness of the
H$\alpha$ filaments in the Perseus cluster suggests that the
intracluster medium may not be highly turbulent (Fabian et al 2003),
although gas flows $\sim 100\mathrm{s}\;\mathrm{km}\;\mathrm{s}^{-1}$
are inevitable. It will be useful to include resonance scattering
effects when examining higher resolution spectra from the
\emph{XMM-Newton} Reflection Grating Spectrometers, or from a future
X-ray bolometer mission.

\subsection{Photoelectric absorption}
Photoelectric absorption affects the spectra emitted from the
innermost regions of clusters of galaxies, as is indicated by the
small scale variations in NGC~4696 (Crawford et al 2005). We find
evidence for larger scale absorbing material in Centaurus, providing
calibration uncertainties are not important. If this absorbing
material can preferentially act on the radiation from cool material,
it has the possibility of disguising the rate of cooling in the
cluster (Fabian et al 2001).

\subsection{Off centre abundance peaks}
Our results show that accounting for projection, resonance scattering
and absorption does not remove the abundance drop seen in the centre
of the Centaurus cluster. Unfortunately this result is not as clear
cut as we would like.

We have shown that when fitting projected spectra, allowing for extra
absorption, and using three temperature components, it is possible to
lose the drop in the central metallicity (Fabian et al 2005). The
uncertainty on the innermost gas abundance does however, increase
substantially, which means the drop may or may not exist.

\begin{figure}
  \includegraphics[width=\columnwidth]{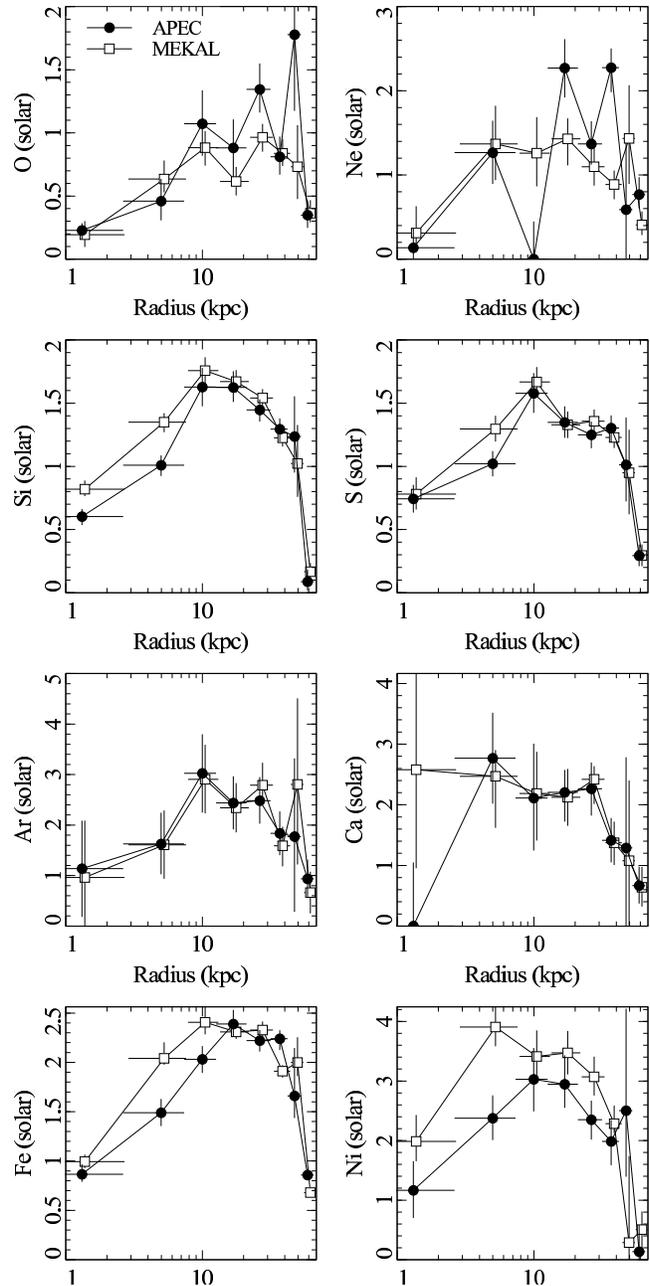}
  \caption{Best fitting metallicities for the western half of the
    cluster when fitting models including projection, using two
    temperature components in each annulus. \textsc{apec} and
    \textsc{mekal} models are compared, with the \textsc{mekal}
    results slightly offset in radius for clarity.}
  \label{fig:2t}
\end{figure}

We have also tried fitting models accounting for projection, but using
two temperature components in each shell. In the fits the O, Ne, Mg,
Si, S, Ar, Ca, Fe and Ni metallicities are allowed to vary, while the
others are fixed at solar. Both of the temperature components are
constrained to have the same metallicites. Fig.~\ref{fig:2t} shows the
results using the \textsc{apec} and \textsc{mekal} (Mewe, Gronenschild
\& van den Oord 1985; Kaastra 1992; Liedahl, Osterheld \& Goldstein
1995) spectral models for the components. It should be noted that the
\textsc{mekal} results used the updated version of the \textsc{mekal}
code which is available in \textsc{spex} (Kaastra 2000). We discuss
this further in Sanders \& Fabian (in preparation), where we will also
examine the relative abundances in more depth, in the context of
models of enrichment.

The plots show that there are declines in metallicity in many
elements, even with this fairly complex model. The declines are not as
deep as for single temperature projected models, however. If there is
additional absorption within the central region (as we may have
detected), or additional temperature components which do not have
spherical symmetry, then this could remove the drop. We have also
tried simple single-temperature models, but including a nonthermal
component in the innermost bins, but this was insufficient.

Although the drop in Centaurus may be removed with a complex enough
spectral model, it is more difficult to remove the decline seen in the
Perseus cluster, for which we have done detailed spatial mapping
(Sanders et al 2004; Sanders, Fabian \& Dunn 2005). This drop has a
large spatial extent, and the distribution of the metals within the
ICM appear to have a great deal of structure. It may be possible that
the inclusion of a range of detected multiphase gas (Fabian et al
2006), and nonthermal emission (Sanders et al 2005), may help remove
the drop, but our initial results indicate that they do not. The
relative amount of cool gas in the Perseus cluster is much smaller
than Centaurus, compared to the hotter gas.  Unfortunately it is
impossible to properly account for projection in Perseus due to the
complex nature of the cluster core. We therefore conclude that there
is much scope for further research into the distribution of the metals
found in cluster cores.

\section*{Model availability}
The model is available for download from
\url{http://www-xray.ast.cam.ac.uk/papers/reson/}

\section*{Acknowledgements}
We would like to thank R. Smith for help in using \textsc{aped} and
\textsc{apec}. ACF acknowledges the support of the Royal Society. 

\appendix
\section{Scattering into the line-of-sight}
\label{appen:scat}
Resonant scattering is not isotropic for $j=0$ and $\Delta j=1$
transitions but has a Rayleigh (dipole) phase function (Hamilton
1947), with the form
\begin{equation}
  \psi(\theta) = \frac{3}{4} \left( 1 + \mathrm{cos}^2 \theta \right),
\end{equation}
which is normalised to $4 \pi$ when integrated over all solid angle.
Other resonance scattering transitions can be isotropic in nature, or
a combination of dipole and isotropic.

We assume a Rayleigh phase function in our calculations, which is true
for many important resonance lines. The choice of phase function,
however, makes little difference to the distribution of the surface
brightness on the sky (Sazonov, Churazov \& Sunyaev 2002), as shown in
our simulation below.

If we assume the photons being scattered are travelling in radial
directions before scattering, then the fraction of photons scattered
from a shell of radius $R_1$ to $R_2$ into an annulus on the sky of
radius $A_1$ to $A_2$ is
\begin{multline}
  f_\mathrm{scat}(R_1,R_2,A_1,A_2)=
  \frac{1}{\frac{4}{3} \pi (R_2^3 - R_1^3)} \\
  \frac{1}{4\pi} \int_{r=R_1}^{R_2} \int_{\phi=0}^{2 \pi}
  \left[
  \int_{\theta = \mathrm{sin}^{-1} \frac{A_1}{r}}^
  {\mathrm{sin}^{-1} \frac{A_2}{r}}
    \frac{3}{4} \left(1 + \mathrm{cos}^2 \theta \right) \mathrm{sin} \theta
    \: \mathrm{d}\theta
  + \right.\\
  \left. 
  \int_{\theta = \pi - \mathrm{sin}^{-1} \frac{A_2}{r}}^
  {\pi - \mathrm{sin}^{-1} \frac{A_1}{r}}
    \frac{3}{4} \left(1 + \mathrm{cos}^2 \theta \right) \mathrm{sin} \theta
    \: \mathrm{d}\theta
  \right]
  \: \mathrm{d}\phi \: r^2 \: \mathrm{d}r,
\end{multline}
assuming that equal photons are scattered by equal volume in the
scattering shell. When integrated this becomes
\begin{multline}
  f_\mathrm{scat}(R_1,R_2,A_1,A_2)=\\
  \frac{1}{4 (R_1^3 - R_2^3)} \left[
    (4 R_1^2 - 7 A_1^2) \sqrt{R_1^2-A_1^2}
    + (7 A_1^2 -4 R_2^2) \sqrt{R_2^2 - A_1^2} \right.\\
   + (7 A_2^2 - 4 R_1^2) \sqrt{R_1^2 - A_2^2}
  + (4 R_2^2 - 7 A_2^2) \sqrt{R_2^2 - A_2^2}\\
  + 3 A_1^3 \left( \mathrm{sin}^{-1}\frac{A_1}{R_2} -
    \mathrm{sin}^{-1}\frac{A_1}{R_1} \right)
  \left. 
  + 3 A_2^3 \left( \mathrm{sin}^{-1}\frac{A_2}{R_1}
  -  \mathrm{sin}^{-1}\frac{A_2}{R_2} \right) \right]. 
\label{eqn:scat}
\end{multline}

\begin{figure}
  \includegraphics[width=\columnwidth]{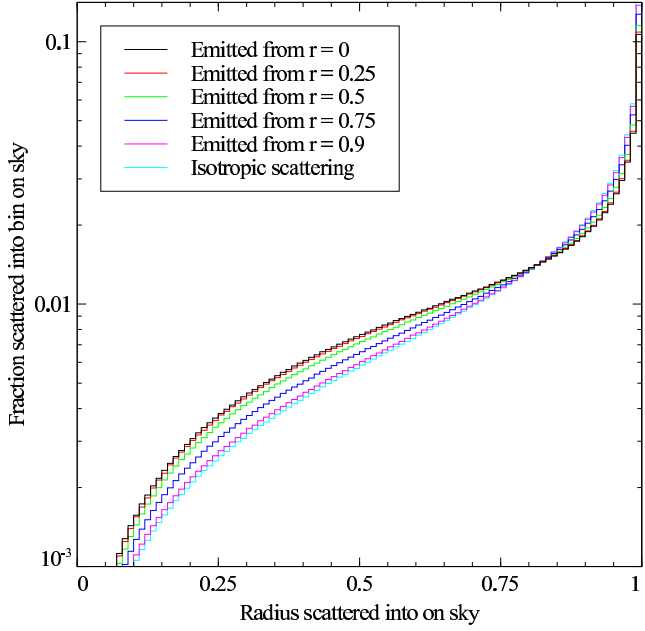}
  \caption{Distribution of light on the sky (in 100 bins) for photons
    randomly emitted isotropically in a system from fixed radii, $r$,
    but scattered by a thin shell at $r=1$. All of these results
    assume Rayleigh scattering, except for the labelled isotropic
    case. As the emitter moves closer to the scatterer, the dipole
    scattering becomes more like isotropic scattering. Note that most
    of the flux is scattered beyond $r=0.75$ in all cases.}
  \label{fig:phasescat}
\end{figure}

If the scatterer is close to the the source of the photons then this
calculation is not accurate. To examine the magnitude of this effect,
we have performed Monte Carlo simulations of individual photons. We
simulated sets of $10^8$ photons emitted at five different radii in
cluster, but scattered at the same radius, for both dipole and
isotropic scattering. Fig.~\ref{fig:phasescat} plots the average
distribution on the sky of the scattered radiation from these
simulations. The results from Equation~\ref{eqn:scat} are exactly the
same as those from the simulation when the light is emitted from $r=0$
with Rayleigh scattering.

Isotropic scattering shows no change in the distribution as the
emitter moves closer to the scatterer. We therefore only plot a single
line for that phase function. For dipole scattering, as the emitter
approaches the scatterer, the distribution becomes more like an
isotropic scatterer.  There is little change in the shape of the
distribution until the emitter is 3/4 of the radius of the scatterer.
As they get closer, the majority of the flux is scattered towards
slightly smaller radii. Note that the magnitude of the effect on the
sky is small. It is magnified by the logarithmic scale used in the
plot, which is necessary to show it.

As the effect of the emitter moving towards the scatterer is
relatively small, and the choice of phase function is small too (also
see Sazonov et al 2002), we assume Equation~\ref{eqn:scat} in our
calculations.


\begin{thebibliography}{}
\bibitem{acton78} Acton L.W., 1978, ApJ, 225, 1069

\bibitem{allen94} Allen S.W., Fabian A.C., 1994, MNRAS, 269, 409

\bibitem{arnaud96} Arnaud K.A., 1996, Astronomical Data Analysis
  Software and Systems V, eds. Jacoby G. and Barnes J., p17, ASP Conf. 
  Series volume 101

\bibitem{balucinskachurch92} Balucinska-Church M., McCammon D., 1992,
  ApJ, 400, 699

\bibitem{buote00} Buote D.A., 2000, ApJ, 539, 172

\bibitem{cash79} Cash W., 1979, ApJ, 228, 939

\bibitem{churazov99} Churazov E., Gilfanov M., Forman W., Jones C.,
  1999, ApJ, 520, 105

\bibitem{churazov03} Churazov E., Forman W., Jones C., B\"ohringer H.,
  2003, ApJ, 590, 225

\bibitem{churazov04} Churazov E., Forman W., Jones C., Sunyaev R.,
  B\"ohringer H., 2004, MNRAS, 347, 29
  
\bibitem{crawford05} Crawford C.S., Hatch N.A., Fabian A.C., Sanders
  J.S., 2005, MNRAS, 363, 216

\bibitem{degrandi01} De Grandi S., Molendi S., 2001, ApJ, 551, 153

\bibitem{dupke01} Dupke R., Bregman J.S., 2001, ApJ, 562, 266

\bibitem{edge91} Edge A.C., Stewart G.C., 1991, MNRAS, 252, 414

\bibitem{fabian94} Fabian A.C., 1994, A\&AR, 32, 277

\bibitem{fabian01} Fabian A.C., Mushotzky R.F., Nulsen P.E.J.,
  Peterson J.R., 2001, MNRAS, 321, L20

\bibitem{fabian03} Fabian A.C., 2003, MNRAS, 344, L27

\bibitem{fabian03b} Fabian A.C., Sanders J.S., Crawford C.S.,
  Conselice C.J., Gallagher III J.S., Wyse R.F.G., 2003, MNRAS, 344,
  L48

\bibitem{fabian05} Fabian A.C. Sanders J.S., Taylor G.B., Allen S.W.,
  2005, MNRAS, 360, L20

\bibitem{fabian06} Fabian A.C., Sanders J.S., Taylor G.B., Allen S.W.,
  Crawford C.S., Johnstone R.M., Iwasawa K., 2006, MNRAS, in press,
  astro-ph/0510476

\bibitem{fukazawa94} Fukazawa Y., Ohashi T., Fabian A.C., Canizares
  C.R., Ikebe Y., Makishima K., Mushotzky R.F., Yamashita K., 1994,
  PASJ, 46, L55

\bibitem{furasho01} Furasho T., et al., 2001, PASJ, 53, 421

\bibitem{gastaldello04} Gastaldello F., Molendi S., 2004, ApJ, 600,
  670

\bibitem{gilfanov87} Gilfanov M.R., Syunyaev R.A., Churazov E.M.,
  1987, Soviet Astr.  Lett., 13, 3

\bibitem{grevsauv98} Grevesse N., Sauval A.J., 1998, Space Sci. Rev.,
  85, 161

\bibitem{hamilton47} Hamilton D.R., 1947, ApJ, 106, 457

\bibitem{johnstone02} Johnstone R.M., Allen S.W., Fabian A.C., Sanders
  J.S., 2002, MNRAS, 336, 299
  
\bibitem{jones02} Jones C., Forman W., Vikhlinin A., Markevitch M.,
  David L., Warmflash A., Murray S., Nulsen P.E.J. 2002, ApJ, 567,
  L115

\bibitem{kaastra92} Kaastra J.S., 1992, An X-Ray Spectral Code for
  Optically Thin Plasmas (Internal SRON-Leiden Report, updated version
  2.0)

\bibitem{kaastra00} Kaastra J.S., 2000, in M.A. Bautista, T.R.
  Kallman, A.K. Pradhan, eds, Atomic Data Needs for X-ray Astronomy,
  \url{http://heasarc.gsfc.nasa.gov/docs/heasarc/atomic/}, 161

\bibitem{liedahl95} Liedahl D.A., Osterheld A.L., Goldstein W.H.,
  1995, ApJ, 438, L115

\bibitem{mathews01} Mathews W.G., Buote D.A., Brighenti F., 2001, ApJ,
  550, L31

\bibitem{mathews03} Mathews W.G., Brighenti F., Buote D.A., Lewis
  A.D., 2003, ApJ, 596, 159

\bibitem{mazzotta98} Mazzotta P., Mazzitelli G., Colafrancesco S.,
  Vittorio N., 1998, A\&AS, 133, 403        

\bibitem{mewe85} Mewe R., Gronenschild E.H.B.M., van den Oord
  G.H.J., 1985, A\&AS, 62, 197
  
\bibitem{morris03} Morris R.G., Fabian A.C., 2003, MNRAS, 338, 824

\bibitem{sanders02} Sanders J.S., Fabian A.C., 2002, MNRAS, 331, 273

\bibitem{sanders04} Sanders J.S., Fabian A.C., Allen S.W., Schmidt
  R.W., 2004, MNRAS, 349, 952

\bibitem{sanders05} Sanders J.S., Fabian A.C., Dunn R.J.H., 2005,
  MNRAS, 360, 133

\bibitem{sazonov02} Sazonov S.Y., Churazov E.M., Sunyaev R.A., 2002,
  MNRAS, 333, 191

\bibitem{schmidt02} Schmidt R.W., Fabian A.C., Sanders J.S., 2002,
  MNRAS, 337, 71

\bibitem{shigeyama98} Shigeyama T., 1998, ApJ, 497, 587

\bibitem{smith01a} Smith R.K., Brickhouse N.S., Liedahl D.A., Raymond
  J.C., 2001a, Spectroscopic Challenges of Photoionized Plasmas, ASP
  Conference Series Vol. 247, eds Gary Ferland and Daniel Wolf Savin. 
  San Francisco: Astronomical Society of the Pacific, p161
  
\bibitem{smith01b} Smith R.K., Brickhouse N.S., Liedahl D.A., Raymond
  J.C., 2001b, Astrophysical Journal, 556, L91

\bibitem{townsley02a} Townsley L.K., Broos P.S., Chartas G.,
  Moskalenko E., Nousek J.A., Pavolv G.G., 2002, Nuc. Instr. and
  Meth. in Phys. Res. A, 486, 716

\bibitem{townsley02b} Townsley L.K., Broos P.S., Nousek J.A., Garmire
  G.P., 2002, Nuc. Instr. and Meth. in Phys. Res. A, 486, 751

\end{thebibliography}
\end{document}